\documentclass[preprint,showkeys,secnumarabic,amsfonts,showpacs,amsmath,amssymb]{revtex4}
\usepackage{epsfig}

\usepackage{euscript,graphicx,amssymb,subfigure}

\usepackage{amsfonts}
\usepackage{amsmath}
\usepackage{amssymb}
\usepackage{graphicx}
\usepackage{euscript}
\usepackage{color}
\usepackage{subfigure}

\def\Box{\hbox{$\rlap{$\sqcup$}\sqcap$}}

\begin{document}

\title{Dynamical stability in scalar–-tensor cosmology}

\author{H. Farajollahi$^{1,2}$}
\email{hosseinf@guilan.ac.ir}
\author{A. Shahabi$^{1}$} \author{A. Salehi$^{1}$}

\affiliation{$^{1}$Department of Physics, University of Guilan, Rasht, Iran}
\affiliation{$^2$ School of Physics, University of New South Wales, Sydney, NSW, 2052, Australia}

\begin{abstract}
We study FRW cosmology for a double scalar–-tensor
theory of gravity where two scalar fields are nonminimally coupled to the geometry. In a framework to study
stability and attractor solutions of the model in the phase space, we constraint the model parameters with the observational data. For an accelerating universe, the model behaves like quintom dark energy models and predicts a transition from quintessence era to phantom era.
\end{abstract}

\maketitle

\section{Introduction}\label{s:intro}

The acceleration expansion of the universe is supported by observations of high redshift type Ia
supernovae, the surveys of clusters of galaxies \cite{Reiss}\cite{Bennet}\cite{Pope}\cite{Riess2}, Sloan digital sky survey ({\bf
SDSS})~\cite{Abazajian} and Chandra X--ray observatory~\cite{Allen}. In addition, Cosmic Microwave Background (CMB)
anisotropies observation \cite{Bennett} exhibit that the universe flatness \cite{Spergel}. The observations determines, with high precisions, the basic cosmological parameters and strongly indicates that the universe is dominated by a smoothly and slowly
varying dark energy (DE). A dynamical equation of state ( EoS) parameter can be considered as an effective parameter to explain the acceleration of the unvierse \cite{Seljak}\cite{Setare}. The parameter is connected directly to the universe energy density and indicate the expansion of the universe \cite{Seljak}\cite{Setare}. In particular, a proposal to explain the recent observations is the quintom dark energy, constructed by two scalar fields, and its EoS parameter crosses the phantom divide line \cite{quintom}\cite{quintom1}\cite{quintom2}\cite{quintom3}\cite{quintom4}.

Motivated from string theories, the scalar-tensor models provide the simplest model-independent description of unification theories which predict couplings between scalars and curvature.
 They have assumed a prominent role since any unification scheme, such as supergravity, in the weak
energy limit, or cosmological models of inflation such as chaotic inflation, seem to be supported by them \cite{capelo}. In addition, they have been employed to study the current acceleration of the universe \cite{Sahoo}\cite{Capozziello}\cite{Faraoni}\cite{Nojiri3}\cite{Setare1}\cite{Carr}.

In this paper, we study the detailed dynamics of the double scalar–-tensor cosmological models. Since the major difficulty in cosmological models is the nonlinearity of the field equations and thus limitation in obtaining the exact solutions, we investigate the asymptotic behaviour
of the model, which provides the relevant features to be compared with the current physical data available
for the universe. In this context, the perturbation methods, especially linear stability analysis which have been used to study the qualitative analysis of the
equations and of the long term behaviour of the solutions are being proposed in this work \cite{stability}\cite{stability1}\cite{stability2}\cite{stability3}\cite{stability4}\cite{stability5}\cite{Chiang}. Sec. two is devoted to a detailed formulation of the model. In Sec. three, we obtain the autonomous equations of the model and the late time attractor solutions by using the phase plane analysis in addition to best fitting the model parameters. We also examine the behavior of the EoS parameter of the model which predict a transition from quintessence era to phantom era in future. In Sec. four, we present summary and remarks.

\section{The model}

A general action in four dimensions, where gravity is nonminimally coupled to two scalar fields with no interaction between fields, is given by \cite{capelo}
\begin{eqnarray}\label{ac1}
S&=&\int[F(\varphi)R+G(\psi)R-\frac{1}{2}\varphi_{;\mu}\varphi^{;\mu}+V(\varphi)\\ \nonumber &-&\frac{1}{2}\psi_{;\mu}\psi^{;\mu}+W(\psi)]
\sqrt{-g}d^{4}x,
\end{eqnarray}
where the functions $F(\varphi)$, $V(\varphi)$, $G(\psi)$, and $W(\psi)$ are not specified. By using the transformations $\phi=F(\varphi)$ and $\omega(\phi)=\frac{F(\varphi)}{2dF(\varphi)/d\varphi}$, the Brans-Dicke action can be simply recovered. Furthermore, in our units, the standard Newton coupling is regained in the limit $F(\varphi)+G(\psi)\longrightarrow -\frac{1}{2}$.
By varying the action with respect to the metric $g_{\mu\nu}$, the field equations can be derived:
\begin{eqnarray}\label{ac2}
[F(\varphi)+G(\psi)](R_{\mu\nu}-\frac{1}{2}g_{\mu\nu}R)=T_{\mu\nu}^{(\varphi)}+T_{\mu\nu}^{(\psi)}
\end{eqnarray}
where the effective stress-energy tensors for the scalar fields $\varphi$ and $\psi$ are:
\begin{eqnarray}
T_{\mu\nu}^{(\varphi)}&=&\frac{1}{2}\varphi_{;\mu}\varphi_{;\nu}-\frac{1}{4}g_{\mu\nu}\varphi_{;\alpha}\varphi^{;\alpha}+\frac{1}{2}g_{\mu\nu}V(\varphi) \nonumber\\
&-&g_{\mu\nu}\Box F(\varphi)+F(\varphi)_{;\mu\nu},\\
T_{\mu\nu}^{(\psi)}&=&\frac{1}{2}\psi_{;\mu}\psi_{;\nu}-\frac{1}{4}g_{\mu\nu}\psi_{;\alpha}\psi^{;\alpha}+\frac{1}{2}g_{\mu\nu}W(\psi) \nonumber\\
&-&g_{\mu\nu}\Box G(\psi)+G(\psi)_{;\mu\nu}.
\end{eqnarray}
The variations with respect to $\varphi$ and $\psi$ give the klein-Gordan equations
\begin{eqnarray}
&&\Box \varphi+R(\frac{dF}{d\varphi})+\frac{dV}{d\varphi}=0,\\
&&\Box \psi+R(\frac{dG}{d\psi})+\frac{dW}{d\psi}=0.
\end{eqnarray}
Let us now take into account a FRW metric of the form
\begin{eqnarray}
ds^{2}=dt^{2}-a^{2}[\frac{dr^{2}}{1-kr^{2}}+r^{2}d\Omega^{2}].\nonumber
\end{eqnarray}
The field equations then become
\begin{eqnarray}
(F&+&G)(2\frac{\ddot{a}}{a}+(\frac{\dot{a}}{a})^{2}+\frac{k}{a}^{2})=-2\frac{\dot{a}}{a}(\dot{\varphi}\frac{dF}{d\varphi}+\dot{\psi}\frac{dG}{d\psi})\nonumber \\ &-&(\dot{\varphi}^{2}\frac{d^{2}F}{d\varphi^{2}}+\ddot{\varphi}\frac{dF}{d\varphi}+\dot{\psi}^{2}\frac{d^{2}G}{d\psi^{2}}+\ddot{\psi}\frac{dG}{d\psi})\nonumber \\
&-&\frac{1}{2}(\frac{1}{2}\dot{\varphi}^{2}+\dot{\psi}^{2}-V-W),\label{ac9}
\end{eqnarray}
\begin{eqnarray}
6(F&+&G)(\frac{\dot{a}}{a})^{2}=-6\frac{\dot{a}}{a}(\dot{\varphi}\frac{dF}{d\varphi}+\dot{\psi}\frac{dG}{d\psi})
\nonumber\\ &+&6\frac{k}{a}^{2}(F+G)+\frac{1}{2}(\dot{\varphi}^{2}+\dot{\psi}^{2})+V+W, \label{ac10}
\end{eqnarray}
\begin{eqnarray}
&&\ddot{\varphi}+3\frac{\dot{a}}{a}=6(\frac{\ddot{a}}{a}+(\frac{\dot{a}}{a})^{2}+\frac{k}{a}^{2})\frac{dF}{d\varphi}
-\frac{dV}{d\varphi},\label{ac11}\\
&&\ddot{\psi}+3\frac{\dot{a}}{a}\dot{\psi}=6(\frac{\ddot{a}}{a}+(\frac{\dot{a}}{a})^{2}+\frac{k}{a}^{2}))\frac{dG}{d\psi}
-\frac{dW}{d\psi},\label{ac12}
\end{eqnarray}
where equation (\ref{ac10}) is the energy constraint corresponding to the (0,0) component of the Einstein equation.

\section{Stability Analysis and cosmological test}

In this section, in a flat FRW universe we study the structure of the dynamical system via  phase plane analysis,
by introducing the following dimensionless variables,
\begin{eqnarray}\label{ac13}
X=\frac{\dot{F}+\dot{G}}{(F+G)H},\ \
Y=\frac{\dot{\varphi}^{2}+\dot{\psi}^{2}}{12(F+G)H^{2}},\ \ Z=\frac{V+W}{6(F+G)H^{2}}
\end{eqnarray}
and also take $\lambda=\frac{\dot{V}+\dot{W}}{(V+W)H}$. Using equations (\ref{ac9})--(\ref{ac12}), the evolution equations of these variables become,
\begin{eqnarray}
X'&=&\frac{2}{X}-\frac{7}{6}+\frac{Z(1-\lambda)}{X}+\frac{4X}{3}+\frac{7Y}{2}-\frac{7Z}{6} \nonumber\\ &-&\frac{4X^{2}}{3}
Z(1-\lambda)-2YX-\frac{Z X}{3},\label{ac14}
\end{eqnarray}
\begin{eqnarray}
Y'&=&2-\frac{4Y}{X}+\frac{4X}{3}-\frac{14Y}{3}+Z -\frac{2ZY}{X}
+\frac{X^{2}}{3}\nonumber\\ &+&\frac{XY}{3}-4Y^{2}+\frac{Z X}{3}-
\frac{2ZY}{3}
\end{eqnarray}
\begin{eqnarray}
Z'=\frac{-5XZ}{3}-\frac{4Z}{X}-\frac{4Z}{3}-\frac{2Z^{2}(1-\lambda)}{X}-4ZY-\frac{2Z^{2}}{3}+\lambda Z,\label{ac16}
\end{eqnarray}
where prime " $'$ "means derivative with respect to $N \equiv ln (a)$. Also, the Friedmann constraint equation (\ref{ac10}) becomes
\begin{eqnarray}\label{ac17}
X+Y+Z=1.
\end{eqnarray}
In term of the new dynamical variable we also have,
\begin{eqnarray}\label{hdot}
\frac{\dot{H}}{H^2}=\frac{2}{X}-\frac{2}{3}+\frac{Z(1-\lambda)}{X}+\frac{X}{3}+2Y+\frac{Z}{3},
\end{eqnarray}
in which can be used to evaluate the effective EoS and deceleration parameters in terms of the new dynamical variables. Using the constraint (\ref{ac17}), the equations (\ref{ac14})--(\ref{ac16}) now reduce to the following two equations:
\begin{eqnarray}\label{ac19}
X'&=& \frac{2}{X}-\frac{7}{3}+\frac{(1-X-Y)(1-\lambda)}{X}+\frac{5X}{2}\nonumber\\&-& \frac{4X^{2}}{3}+\frac{14Y}{3}
-(1-X-Y)(1-\lambda)-2YX  \nonumber\\ &-&  \frac{(1-X-Y)X}{3},
\end{eqnarray}
\begin{eqnarray}\label{ac20}
Y'&=&2-\frac{4Y}{X}+\frac{4X}{3}-\frac{14Y}{3}+(1-X-Y)(1-\lambda) \nonumber\\ &-&\frac{2(1-X-Y)Y}{X} + \frac{X^{2}}{3}\nonumber\\
\nonumber &+&\frac{XY}{3}-4Y^{2}+\frac{(1-X-Y)X}{3}-\frac{2(1-X-Y)Y}{3}\nonumber\\  &-&\lambda (1-X-Y).
\end{eqnarray}
It is more appropriate to solve the dynamical equations (\ref{ac19})--(\ref{ac20})
than (\ref{ac14})--(\ref{ac16}). In stability analysis, by simultaneously solving $X'=0$ and $Y'=0$, we find the critical points (fixed points) and study
the stability of these points. In the
context of autonomous dynamical systems, these points are exact constant solutions and are often the extreme points of
the orbits and therefore describe the asymptotic behavior of the system. To the first orders in the perturbations, by substituting
linear perturbations $X'\rightarrow X'+\delta X'$, $Y'\rightarrow Y'+\delta Y'$ about the critical points into the two independent equations (\ref{ac19}) and (\ref{ac20}), we yield two eigenvalues $\lambda_{i} (i=1,2)$. For a stable point, we require the real part of all the eigenvalues to be negative. We find eigenvalues for two critical points. The expression for both critical points and eigenvalues in our model depend on the stability parameter $\lambda$, and are highly nonlinear, long and cumbersome. Therefore, one can not easily reveal the nature of critical points. To circumvent this issue in addition to physically motivate the subject, we simultaneously solve the equations by best fitting the stability parameter and initial conditions with the observational data using the $\chi^2$ method.  Next, we solve the equations by best fitting the model with the observational data for distance modulus.

\subsection{Observational constraints }

The distance modulus is defined as, $\mu(z) = 25 + 5\log_{10}d_L(z)$ where the luminosity distance quantity, $d_L(z)$ is
\begin{equation}\label{dl}
d_{L}(z)=(1+z)\int_0^z{-\frac{dz'}{H(z')}}.
 \end{equation}
 By solving the equations, we find $H(z)$ which can be used to evaluate $\mu(z)$. To best fit the model for the parameter $\lambda$ and the initial conditions $Y(0)$, $X(0)$, $H(0)$ with the most recent observational data, the Type Ia supernovea (SNe Ia), we employe the $\chi^2$ distance analysis. We constrain the parameters including the initial conditions by minimizing the $\chi^2$ function given as
\begin{eqnarray}\label{chi2}
 &&\chi^2_{SNe} ( \lambda, X(0), Y(0), H(0)) \nonumber \\
&=& \sum_{i=1}^{557}\frac{[\mu_i^{the}(z_i|\lambda, X(0), Y(0), H(0)) - \mu_i^{obs}]^2}{\sigma_i^2},
\end{eqnarray}
where the sum is over the SNe Ia sample. In the above relation, $\mu_i^{the}$ and $\mu_i^{obs}$ are the distance modulus parameters obtained from model and observation, respectively. Also, the estimated error of the $\mu_i^{obs}$ is $\sigma$. For our model the best fit values for the parameter and initial conditions occur at $\lambda=2.02$, $X(0)=-0.3$, $Y(0)=-0.65$ and $H(0)=0.91$ with $\chi^2_{min}=552.7337761$. Fig. 1) shows the 1-dim and 2-dim likelihood for the parameters $\lambda$ and the Hubble parameter $H(0)$ at the $68.3\%$, $95.4\%$ and $99.7\%$ confidence levels.

\begin{figure}[t]
\includegraphics[scale=.3]{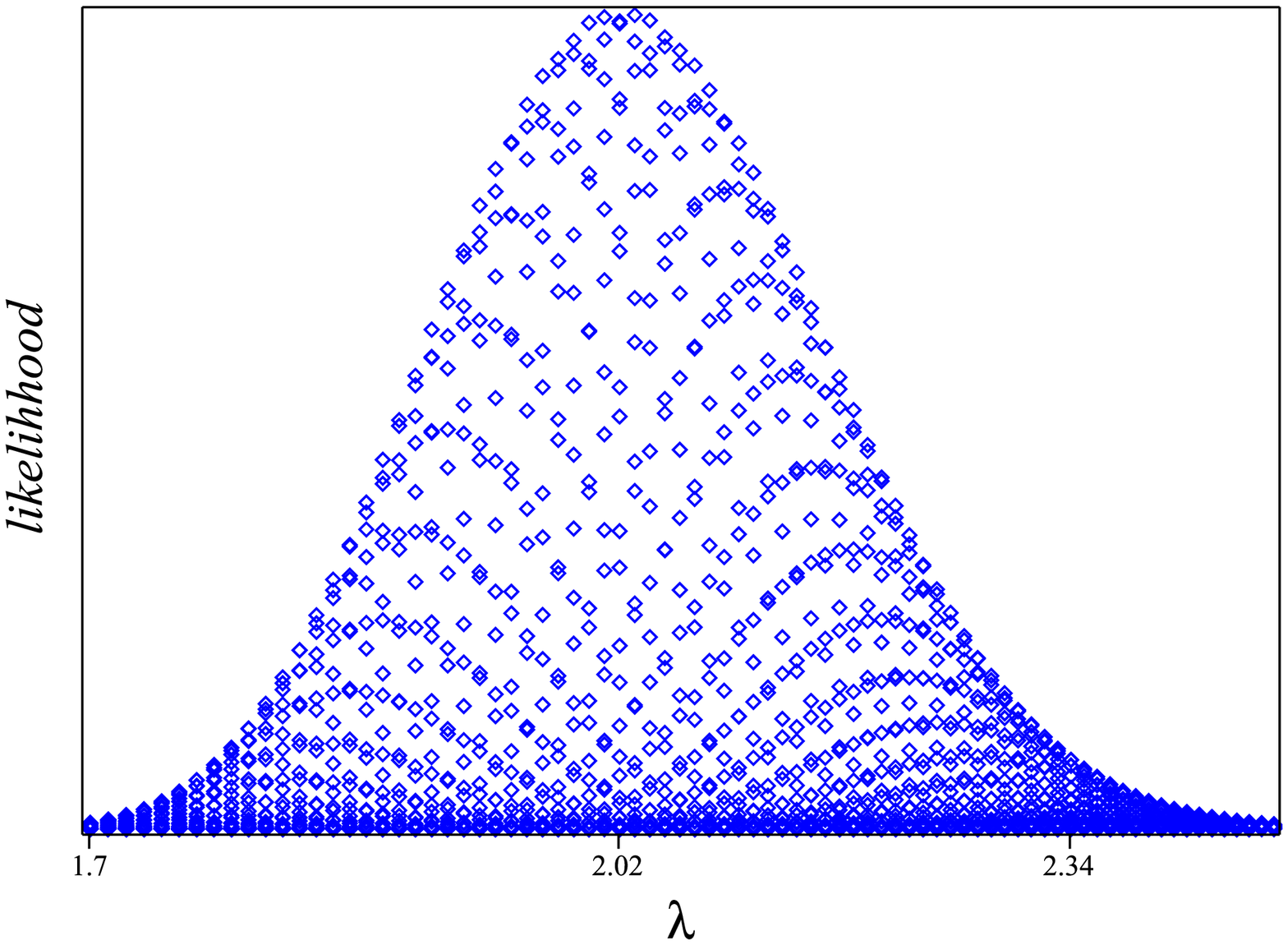}\hspace{0.1 cm}\includegraphics[scale=.3]{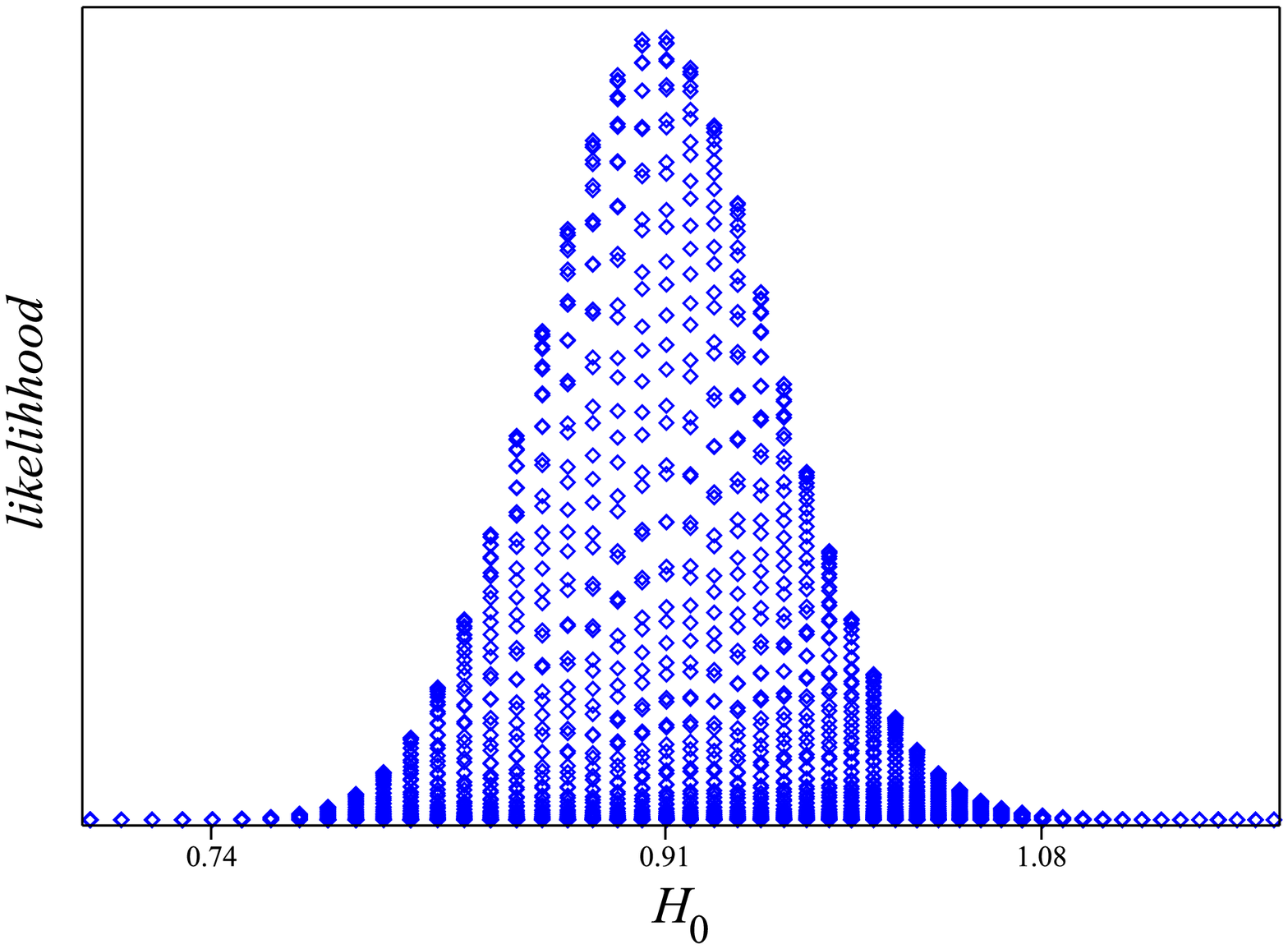}\hspace{0.1 cm}\includegraphics[scale=.3]{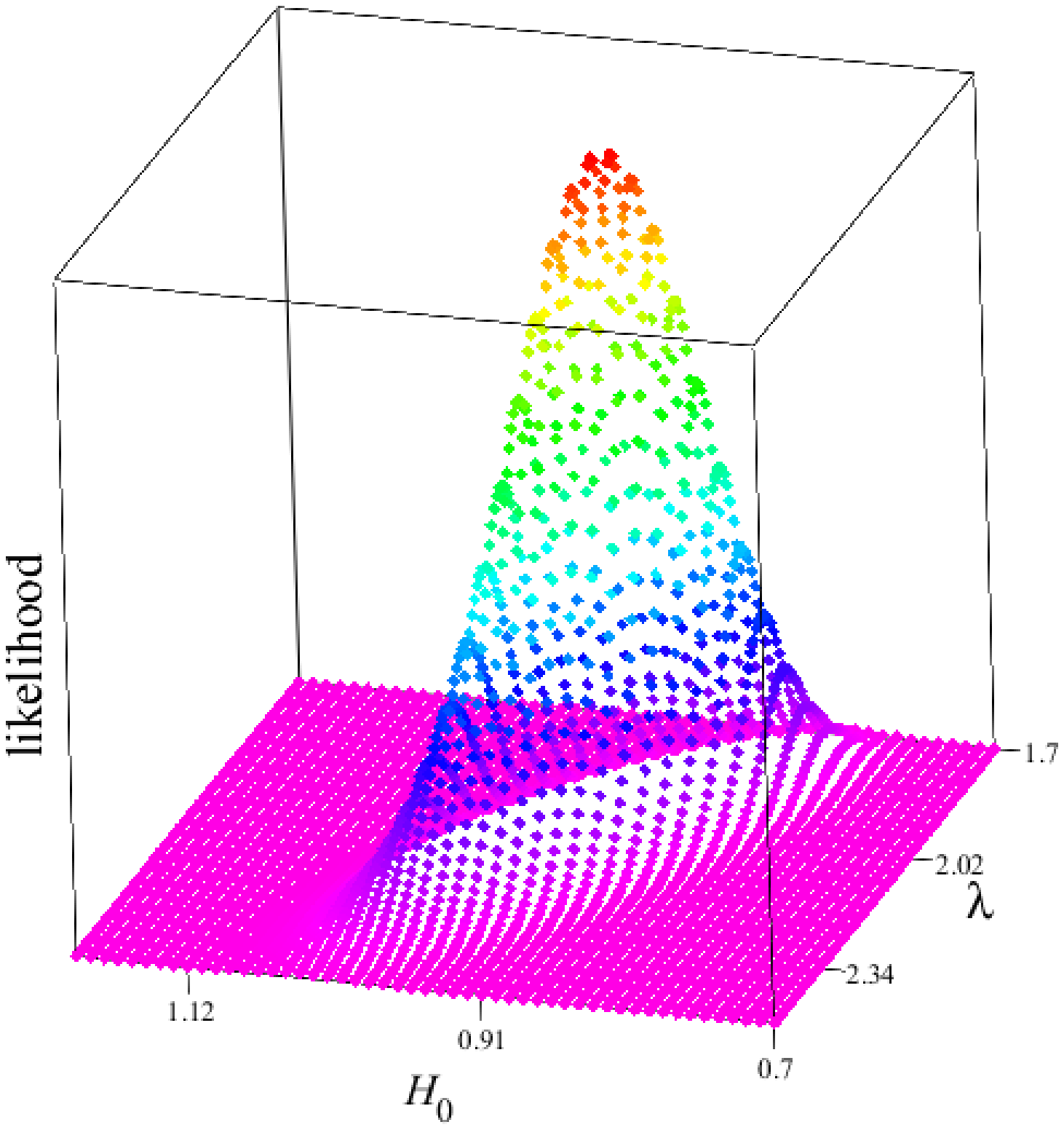}\hspace{0.1 cm}\\
Fig. 1:  The graph of 1-dim and 2-dim likelihood distribution for parameters $\lambda$ and $H(0)$
\end{figure}

In Fig. 2, the distance modulus, $\mu(z)$, in our model is compared with the observational data for the obtained parameters and initial conditions using $\chi^2$ method. As can be seen the best fitted parameters and initial conditions are in good agreement with the observational data. In the following we investigate the stability of the model
with respect to the best fitted model parameter.\\

\begin{figure}[t]
\includegraphics[scale=.4]{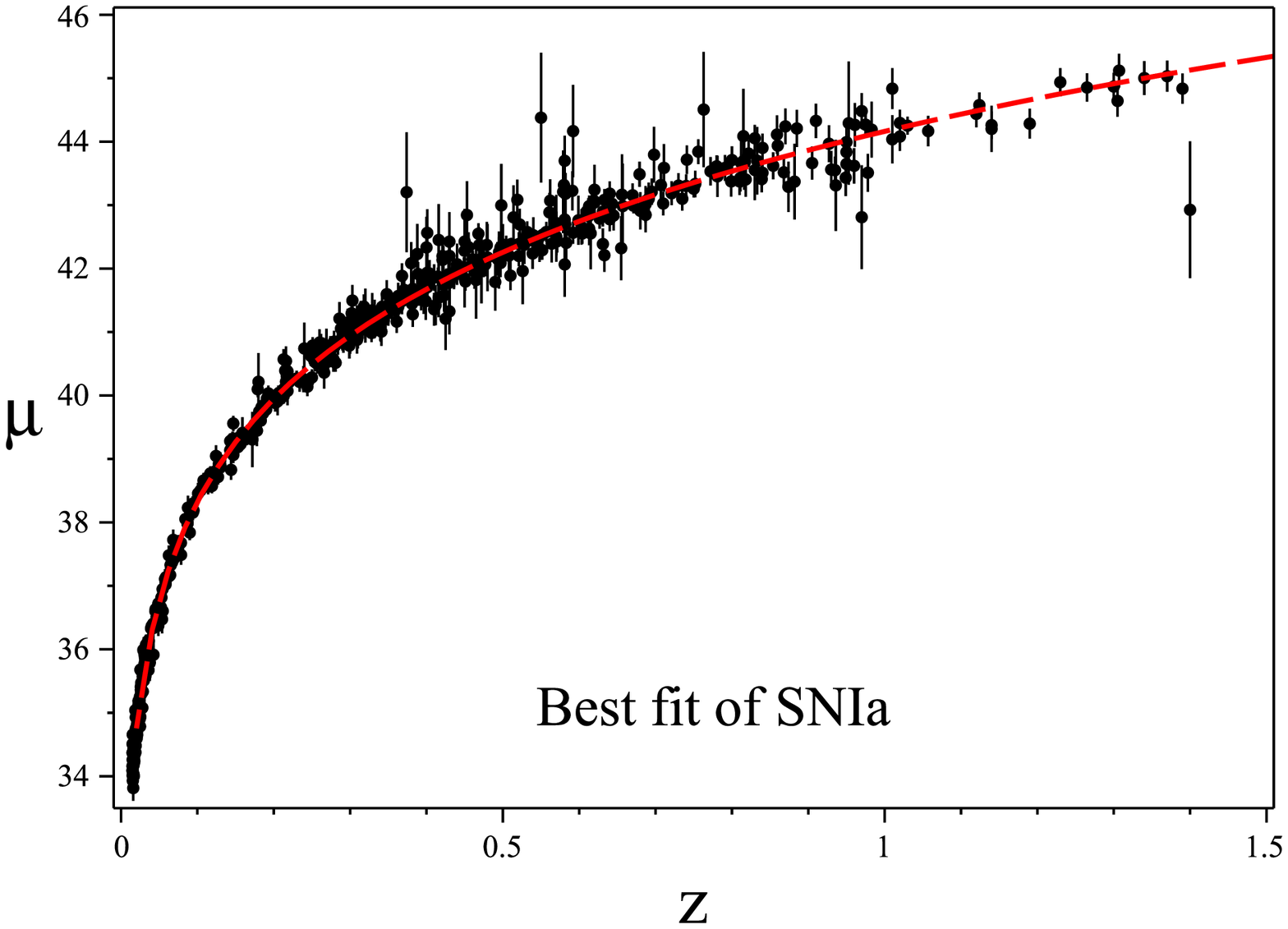}\hspace{0.1 cm}\\
Fig. 2: The best fitted distance modulus $\mu(z)$ plotted as function of redshift
\end{figure}

\subsection{Stability and phase space}

Solving the stability equations for the best fitted model parameter $\lambda$ we find two fixed points with the stability properties given in tables I.

\begin{table}[ht]
\caption{Best fitted critical points} 
\centering 
\begin{tabular}{c c c c c } 
\hline\hline 
Critical point  & $(X, Y)  $  & stability \\  
\hline 
 FP1&(1.72, 0.55) & stable
 \\
 \hline
 FP2 & $(-0.5, 1.26)$ & stable
 \\
\end{tabular}
\label{table:1} 
\end{table}

From the above table we find that both critical points are stable. In Fig. 3, the trajectories entering the stable critical points FP1 and FP2 in the phase plane are illustrated. For the given initial conditions on $X$, $Y$ between -1 and 1, the trajectories (green curves) approaching the stable critical point FP1 in the phase plane are shown. For the initial conditions on $X$, $Y$ less than -1 or greater than 1, then the trajectory shown entering the stable critical point FP2. The best fitted trajectory ( red curve) with the properties given in the previous section enters the stable critical point FP1. This trajectory both fits the model parameter $\lambda$ and the initial conditions for $X$, $Y$ and $H$.

\begin{figure}[t]
\includegraphics[scale=.45]{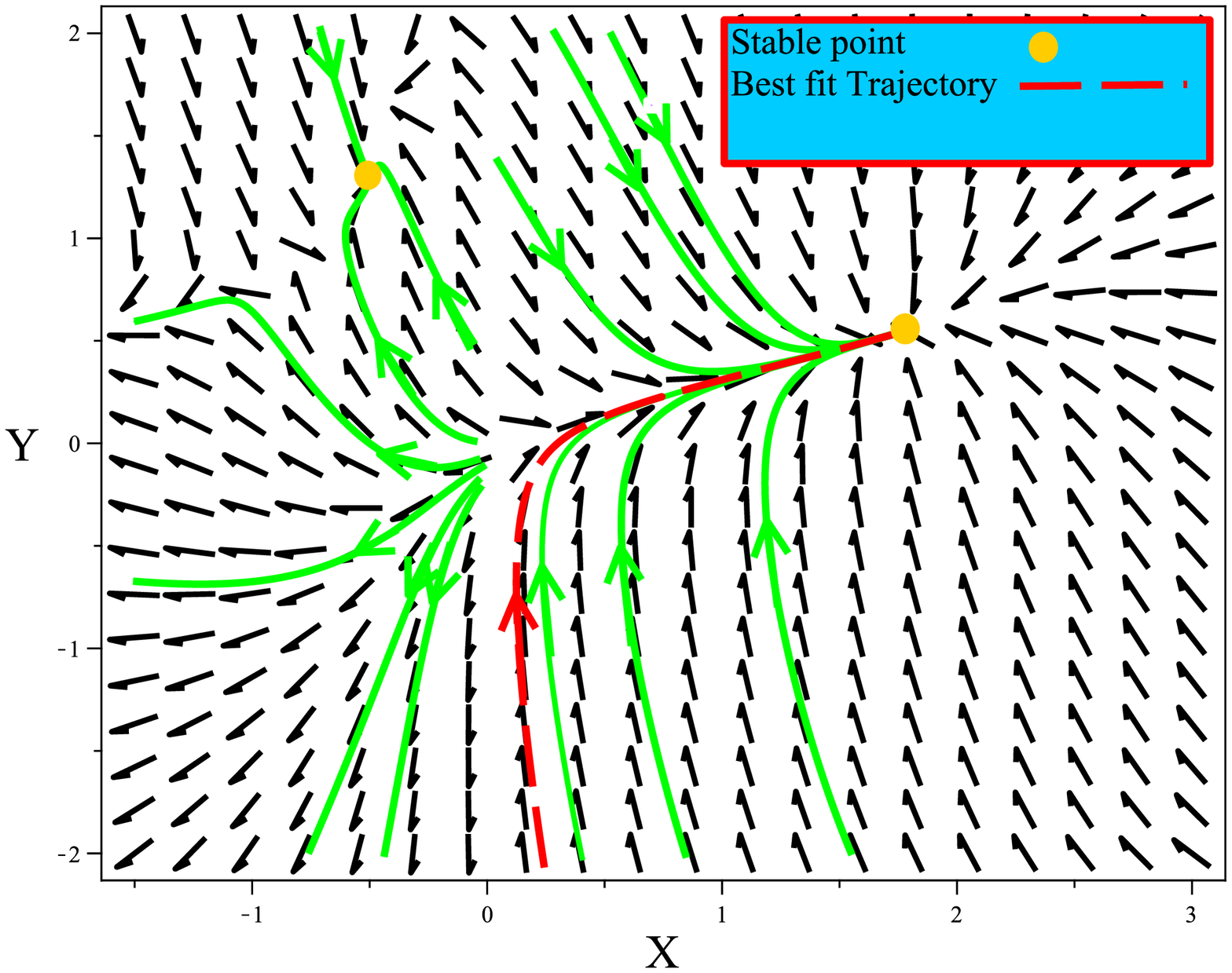}\hspace{0.1 cm}\\
Fig. 3:  The attractor property of the dynamical system in
the phase plane.  The red trajectory approaching the FP1 corresponds to the
best fitted stability parameter and initial conditions.
\end{figure}

\subsection{Phantom crossing}

The EoS parameter is one of the most important cosmic parameters that shows the behavior of the universe and its dynamics. In this work, with stability analysis in addition to constraining the model with the observational data we obtain a better understanding of the cosmological paraments such as EoS parameter. The effective EoS parameter is given by $\omega_{eff}=-1-\frac{2}{3}\frac{\dot{H}}{H^{2}}$ where $\frac{\dot{H}}{H^{2}}$ in terms of best fitted model parameters is given in equation (\ref{hdot}). From Fig. 4, the model does not predict a matter dominated epoch in the higher redshift. However, it exhibits phantom crossing behavior in the future (Fig. 4). The current best fitted effective EoS parameter is $\omega_{eff} \simeq -0.45$. Fig. 4 also illustrate that the phantom crossing occurs in future at $z\simeq -0.4$ and approaches $-1.1$ at $z=-1$ when a big rip occurs.

\begin{figure}[t]
\includegraphics[scale=.45]{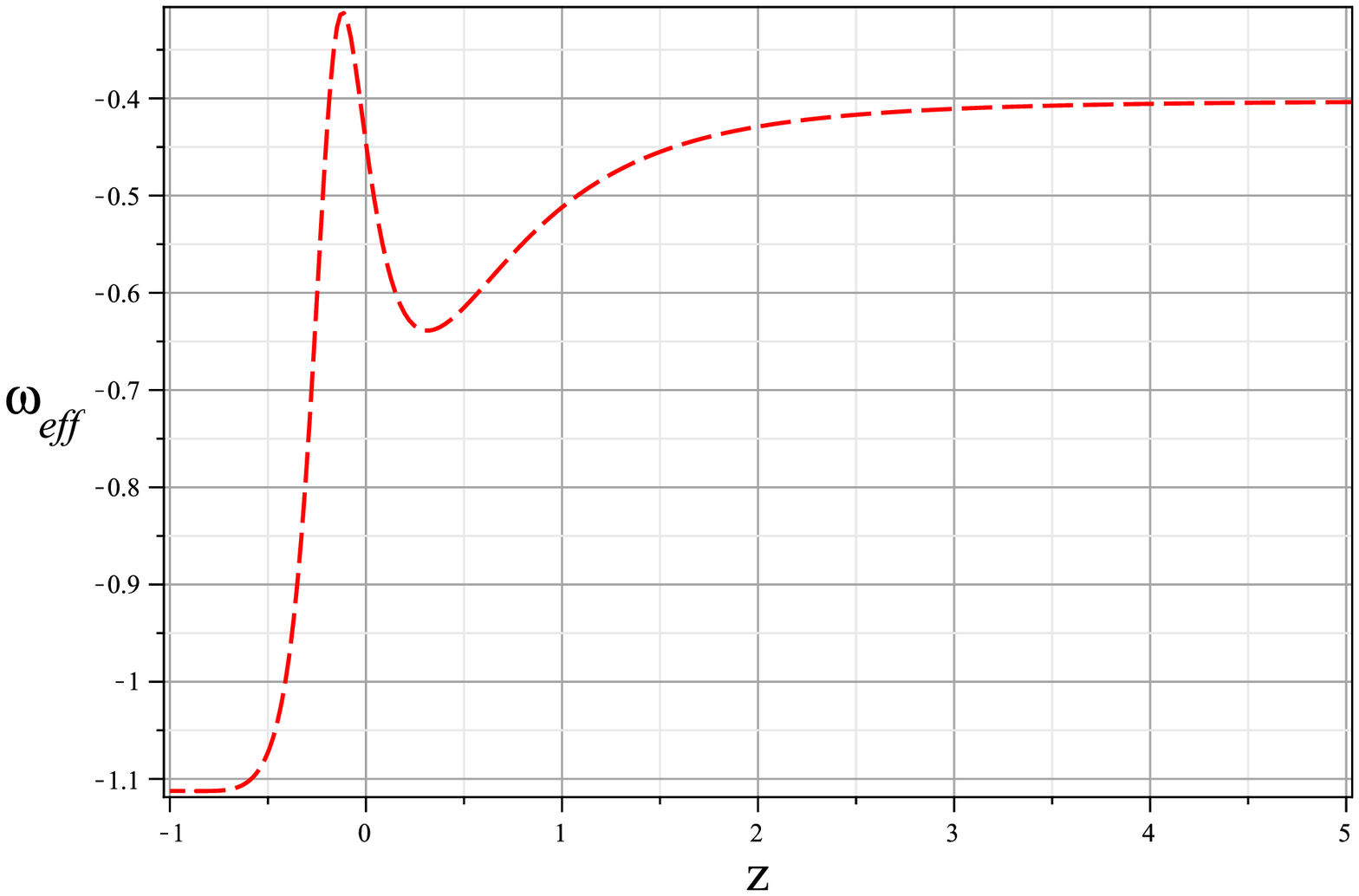}\hspace{0.1 cm}\\
Fig. 4: The best fitted effective EoS parameter, $\omega_{eff}$, as function of redshift. \\
\end{figure}

\section{Summary and Remarks}

This paper is designed to study dynamics of the double scalar–-tensor gravity cosmology by
using the stability analysis and the 2-dimensional phase space of the theory. Different from the work in \cite{stability}, in a remarkable approach in stability analysis, we solve the system of autonomous differential equations by constraining the model parameters and also the initial conditions with the observational data for distance modulus. As a result, the critical points which exhibit states of the universe are observationally verified. From analysis, the two stable critical points in the model are depicted in Figs. 3. With the constrained parameters, the best fitted trajectory approaches only the first stable point.

We then test the model against observational data by calculating the best fitted effective EoS parameter, $\omega_{eff}$ for the model in terms of the stability parameters. The result shows that our model exhibits a quintom behavior; an accelerating universe with a transition from $\omega_{eff}>-1$ (quintessence era) to $\omega_{eff}<-1$ (phantom era)in the future.

\end{document}